\documentclass[conference]{IEEEtran}

\usepackage[english]{babel}

\usepackage{ifluatex}
\ifluatex
  \usepackage{fontspec}
  \usepackage[english]{selnolig}
\fi

\ifluatex
\else

  \usepackage[T1]{fontenc}
  \usepackage[utf8]{inputenc}
\fi

\usepackage{graphicx}

\usepackage{upquote}

\usepackage{booktabs}

\usepackage{paralist}

\usepackage{csquotes}

\usepackage{textcmds}

\RequirePackage[
  babel,
  final,
  expansion=alltext,
  protrusion=alltext-nott]{microtype}

\DisableLigatures{encoding = T1, family = tt* }

\usepackage{url}

\makeatletter
\g@addto@macro{\UrlBreaks}{\UrlOrds}
\makeatother

\usepackage{xcolor}

\usepackage{listings}
\renewcommand{\lstlistingname}{List.}

\usepackage{chngcntr}
\AtBeginDocument{\counterwithout{lstlisting}{section}}

\usepackage{pdfcomment}

\ifluatex

\else

  \SetExpansion
  [ context = sloppy,
    stretch = 30,
    shrink = 60,
    step = 5 ]
  { encoding = {OT1,T1,TS1} }
  { }
\fi

\usepackage{stfloats}
\fnbelowfloat

\usepackage{hyperref}

\hypersetup{hidelinks,
  colorlinks=true,
  allcolors=black,
  pdfstartview=Fit,
  breaklinks=true}

\usepackage[all]{hypcap}

\usepackage[group-four-digits,per-mode=fraction]{siunitx}

\usepackage{amsmath}

\usepackage[capitalise,nameinlink]{cleveref}

\usepackage{iflang}
\IfLanguageName{ngerman}{
  \crefname{table}{Tab.}{Tab.}
  \Crefname{table}{Tabelle}{Tabellen}
  \crefname{figure}{\figurename}{\figurename}
  \Crefname{figure}{Abbildungen}{Abbildungen}
  \crefname{equation}{Gleichung}{Gleichungen}
  \Crefname{equation}{Gleichung}{Gleichungen}
  \crefname{listing}{\lstlistingname}{\lstlistingname}
  \Crefname{listing}{Listing}{Listings}
  \crefname{section}{Abschnitt}{Abschnitte}
  \Crefname{section}{Abschnitt}{Abschnitte}
  \crefname{paragraph}{Abschnitt}{Abschnitte}
  \Crefname{paragraph}{Abschnitt}{Abschnitte}
  \crefname{subparagraph}{Abschnitt}{Abschnitte}
  \Crefname{subparagraph}{Abschnitt}{Abschnitte}
}{
  \crefname{section}{Sec.}{Sec.}
  \Crefname{section}{Section}{Sections}
  \crefname{listing}{\lstlistingname}{\lstlistingname}
  \Crefname{listing}{Listing}{Listings}
}

\ifluatex
\else

  \input glyphtounicode
\fi

\usepackage{verbatim}

\usepackage{tikz}
\usetikzlibrary{arrows,automata, shapes, plotmarks, decorations.pathreplacing}
\usetikzlibrary{decorations.pathreplacing}
\usetikzlibrary{positioning, fit}

\DeclareMathOperator{\hash}{hash}

\begin{document}

\title{Erasing Data from Blockchain Nodes}

\author{\IEEEauthorblockN{Martin Florian, Sophie Beaucamp, Sebastian Henningsen, Björn Scheuermann}
\IEEEauthorblockA{Humboldt Universität zu Berlin / Weizenbaum Institute}
}

\maketitle

\begin{abstract}
  It is a common narrative that blockchains are immutable and so it is technically impossible to erase data stored on them.
  For legal and ethical reasons, however, individuals and organizations might be compelled to erase locally stored
  data, be it encoded on a blockchain or not.
  The common assumption for blockchain networks like Bitcoin is that
  forcing nodes to erase data contained on the blockchain is equal to
  permanently restricting them from
  participating in the system in a full-node role.
  Challenging this belief, in this paper, we propose and demonstrate
  a pragmatic approach towards
  functionality-preserving local erasure~(FPLE).
  FPLE enables full nodes to erase infringing or undesirable data
  while continuing to store and validate most of the blockchain.
  We describe a general FPLE approach for UTXO-based (i.\,e., Bitcoin-like) cryptocurrencies
  and present a lightweight proof-of-concept tool for safely erasing transaction data
  from the local storage of Bitcoin Core nodes.
  Erasing nodes continue to operate in tune with the network
  even when erased transaction outputs become relevant for validating subsequent blocks.
  Using only our basic proof-of-concept implementation,
  we are already able to safely comply with a significantly larger range of erasure
  requests than, to the best of our knowledge, any other full node operator so far.
\end{abstract}

\section{Introduction}
\label{sec:intro}

Ten years since the birth of
Bitcoin~\cite{nakamoto2008bitcoin,tschorsch2016bitcoin},
worldwide enthusiasm for its technological approach is still at a high.
Decentralized networks maintaining append-only ledgers, or
\emph{blockchain} networks, capture the imagination and are heralded as
solutions to decade-old challenges of trust and control.
Blockchain networks depend on a healthy population of \emph{full nodes},
i.\,e. individual peers that store, verify and distribute the full set of data
comprising the blockchain---the history of all past transactions.
And blockchains are commonly advertised as being \emph{immutable}, making it
impossible to change or erase already posted data.
Consequently, potential node operators today face a binary
choice: deploy a full node, contribute to overall network resilience, and store all the data that anyone ever added to the blockchain---or participate as a light client with reduced security and privacy~\cite{gervais2014spvprivacy}.
Blockchains can and do hold arbitrary information in addition to financial transactions, though.
And there is a wide spectrum of legal and ethical reasons why individual node operators
might refuse to store and distribute certain data.
Even one inappropriate transaction might flip the above
decision~\cite{matzutt2018quantitative, beaucamp2018strafbarkeit}.

Is a third way possible?
Can we give potential node operators stronger
options to contribute, keeping the network healthy,
while still complying with local regulation and individual
values?
We support this view and argue that a more thorough discussion of \emph{local erasure}
in existing popular blockchain networks such as Bitcoin is overdue.
We propose the concept of
\emph{functionality-preserving local erasure} (\emph{FPLE})---a
pragmatic solution to the challenge of
locally erasing data that is
potentially consensus-critical in the future while
retaining core full node functions and benefits.
In contrast to previous erasure proposals~\cite{ateniese2017redactable, puddu2017muchain, deuber2019redactable},
FPLE requires no protocol
changes and is fully compatible with existing UTXO-based (i.\,e., Bitcoin-like)
networks without causing forks
or introducing new points of trust.
While FPLE enables only the local erasure of data, we argue that this is a
necessary building block for enabling global erasure without introducing
significant changes to the existing trust model.

As a proof-of-concept, we implemented a prototypic tool for operators
of Bitcoin full nodes.
Our tool enables the erasure of
transaction outputs from the local data stores of \emph{Bitcoin Core} nodes,
while enabling the nodes to stay in sync with the network
even if the erased outputs are later spent.
This does not require any changes to the node software itself or to the Bitcoin protocol.
Our tool is thoroughly tested against current Bitcoin Core versions and
enables us to safely comply with a larger range of erasure requests
than previously possible for node operators.

Our main contributions can be summarized as follows:
\begin{itemize}
  \item A pragmatic and individually deployable
    approach towards functionality-preserving local erasure
    (FPLE) that
    provides solutions for challenges such as the erasure of
    potentially consensus-critical data (\cref{sec:approach}).
  \item A proof-of-concept tool that demonstrates the feasibility
    of applying FPLE to existing Bitcoin nodes (\cref{sec:poc}).
  \item A thorough discussion of legal and non-legal reasons for
    local erasure (\cref{sec:why_erase_}) and
    of potential implications of FPLE when applied to existing networks
    (\cref{sec:discussion}).
\end{itemize}

\section{Related Work}
\label{sec:related_work}

It is no secret that arbitrary data can be included on blockchains---generic non-financial data was included as early as in Bitcoin's genesis block.
Non-financial data storage on blockchains enables innovative new services such as
name services\footnote{\url{https://www.namecoin.info/}},
timestamping\footnote{\url{https://opentimestamps.org/}},
pseudonymous identities~\cite{florian2015sybil} and
non-equivocation logging~\cite{tomescu2017catena}
(to name just a few examples).
Recent results, however, demonstrate
that an uncensorable data storage service like Bitcoin
can also be abused, respectively that some of the
data stored on it might not be universally well looked upon~\cite{matzutt2018quantitative}.
A range of solutions exist for alleviating this conflict.
They can be grouped into the categories: avoiding
the inclusion of unwanted data, allowing the modification (and erasure) of
past blockchain state, and local pruning.
FPLE can be seen as
an improvement to local pruning. Most importantly, we aim at solving the
data erasure challenge node-locally instead of on a global level.

\subsection{Avoiding Unwanted Data}
\label{sub:filtering_data}

In~\cite{matzutt2018thwarting}, Matzutt et al. discuss various approaches for
preventing the insertion of arbitrary, potentially unwanted data onto
cryptocurrency blockchains.
Their proposals include content detectors,
which filter transactions based on heuristics
and knowledge about commonly used data insertion methods,
as well as protocol modifications that would
greatly increase the costs of including arbitrary data.
Approaches along these lines have also surfaced in the non-academic
cryptocurrency community\footnote{
  See, e.\,g.:
  \url{http://comments.gmane.org/gmane.comp.bitcoin.devel/1996}
}.
Approaches for avoiding the insertion of unwanted data depend on global
adoption, for an effective filtering, and in some cases also on protocol
changes when applied to existing networks.
In contrast, FPLE requires only a node-local decision and is in this
way both more practical and enables the incorporation of a wider range of
individual preferences and constraints.
Lastly, as can be seen in related application domains such as
malware detection or digital rights protection via upload filtering,
content-based filtering is never completely circumvention-proof.
Once something "slips through", an erasure possibility again becomes necessary.

When considering data protection as a reason for erasure (cf.
\cref{sub:data_protection}), it also noteworthy that a large body of works
deal with the challenge of providing \emph{anonymity} to blockchain users
(see e.\,g. \cite{conti2018survey} for a recent survey).
However, most transactions in popular systems like Bitcoin do not use any additional means of increasing
anonymity \cite{moser2017anonymous} and are reidentifiable using well-known techniques \cite{conti2018survey}.
Even when strong privacy guarantees can be achieved through technical means,
this provides no solution for cases where identifiable data is posted to the blockchain on purpose, e.\,g., as part of doxing.

\subsection{Redacting the Blockchain}
\label{sub:redacting}

A straightforward approach to globally
erasing previously included data from a blockchain is to
produce a \emph{hard fork}~\cite{ateniese2017redactable}.
Safe hard forks require a strong off-chain
consensus among miners, users and network operators. In public networks with
little central coordination, such as
Bitcoin, such a consensus is notoriously difficult to achieve.
Even more so when compared to the ease of including potentially problematic data
at a high rate.
\emph{Redactable blockchains}~\cite{ateniese2017redactable} have been proposed
as an alternative blockchain design that allows the global erasure of data without
causing hard forks.
They use \emph{chameleon hash functions}~\cite{camenisch2017chameleon} that enable
trusted entities with access to a trapdoor key to calculate hash collisions
and therefore change published data while maintaining the appearance of chain integrity.
Alternative solutions were proposed that deal with resulting trust problems
by employing a voting-like approach~\cite{puddu2017muchain, deuber2019redactable}.
However, \cite{ateniese2017redactable}, \cite{puddu2017muchain} and \cite{deuber2019redactable}
require heavy changes to existing systems, also altering their underlying trust model.
In contrast to their motivation of erasing data globally, we focus on local
erasure without requiring protocol changes.

\subsection{Pruning}
\label{sub:pruning}

\emph{Pruning} is a widely used technique for locally erasing older parts of
a blockchain, mainly with the goal of reducing storage requirements.
While related, our local erasure approach differs in its goal---we erase
individual data chunks instead of the whole history before a certain point---and provides solutions for outstanding challenges such as the pruning of data
potentially relevant for validating future blocks.
The latter challenge is highly relevant in practice as problematic data is
often encoded in unspent but potentially spendable transaction
outputs~\cite{matzutt2018quantitative}.

Alternative blockchain designs such as \cite{chepurnoy2016rollerchain}
propose storing the current global state,
e.\,g., in terms of account balances,
in each block so that older blocks can be more safely pruned.
A per-block cryptographic commitment to the current state is also used in
popular networks such as Ethereum~\cite{wood2014ethereum}.
While potentially making past transactions more easily prunable,
neither of these solutions help in cases where potentially unwanted data can
be reconstructed from the current state, such as when it is encoded in account
addresses or smart contract data.
With FPLE, we explicitly consider the erasure of data that is part of the
\emph{UTXO set}, the equivalent of "state" in UTXO-based systems.

\section{A Pragmatic Approach to Erasure}
\label{sec:approach}

In the following, we present a specific proposal towards
\emph{functionality-preserving local erasure} (FPLE).
We propose FPLE as an extension to existing node software for common
blockchain networks like Bitcoin. FPLE enables individual node operators to
mark chunks of data (e.\,g., transaction outputs) for erasure
without requiring protocol changes,
coordination with other nodes or the introduction of global trust anchors.
In \cref{sec:poc}, we even show how, when targeting Bitcoin, core benefits
of our approach can be reaped without making changes to the actual node software.
For enabling a more concise description,
we focus on erasure
\emph{from}
UTXO-based (i.\,e., Bitcoin-like) blockchains and \emph{of} data stored in
\emph{transaction outputs}.
Transaction outputs are the most heavily used data storage
location in Bitcoin~\cite{matzutt2018quantitative} and
one of the most challenging when it comes to functionality-preserving erasure.
We give an overview over possible data storage locations in
\cref{sub:utxo_model},
after first giving an overview over our general approach.

\subsection{Basic Approach}
\label{sub:nutshell}

With FPLE, chunks of data
marked for erasure are physically erased from storage or garbled,
never again stored in a reconstructable form,
and never shared with other nodes.
Conceptually,
references to erased data are stored in an
\emph{erasure database}, to avoid requesting unwanted data in the future, filtering
it upon renewed receipt and being able to differentiate non-existent data from erased data.

For dealing with possible validation
challenges following the erasure of data chunks,
a set of pragmatic workarounds is proposed.
Namely, we employ the following principles
whenever validation steps depend
on portions of erased data:

\begin{enumerate}
  \item Nodes ignore and never relay unconfirmed transactions whose validation depends on erased data.
  \item If a transaction depending on erased data is included on the blockchain, erasing nodes assume that the
    verification operations relating to the erased data finished positively.
\end{enumerate}

In essence, for the subset of transactions and blocks for which validation
\emph{directly depends} on raw erased data (and \emph{only} for that subset), the erasing node assumes
a mode of validation akin to the \emph{simplified payments verification} (SPV)
paradigm---it trusts that miners are sufficiently incentivised to mine valid blocks.
We focus on this simple and universally applicable approach to prove our point that erasure from blockchain nodes is not an impossibility.
However, we also discuss ideas towards fully eliminating the need for an SPV fallback (in \cref{sub:trustless}).
In the following,
building upon an abstract model of UTXO-based blockchain protocols, we further specify our
proposals for erasure and validation and discuss potential security implications of our design.
In \cref{sec:poc} we then introduce a lightweight
proof-of-concept implementation of FPLE for
\emph{Bitcoin Core}.

\subsection{UTXO-based Blockchain Protocols}
\label{sub:utxo_model}

We introduce our FPLE approach based on an abstract model of UTXO-based blockchain protocols
inspired by the description in~\cite{tschorsch2016bitcoin}.
We focus on the aspects of UTXO-based blockchain protocols that are most
relevant for the storage and erasure
of arbitrary data, namely the defining data structures and their fields,
and the mechanisms of validation (as they might depend on erased data).

\begin{figure}[htpb]
  \centering
  \includegraphics[width=0.8\linewidth]{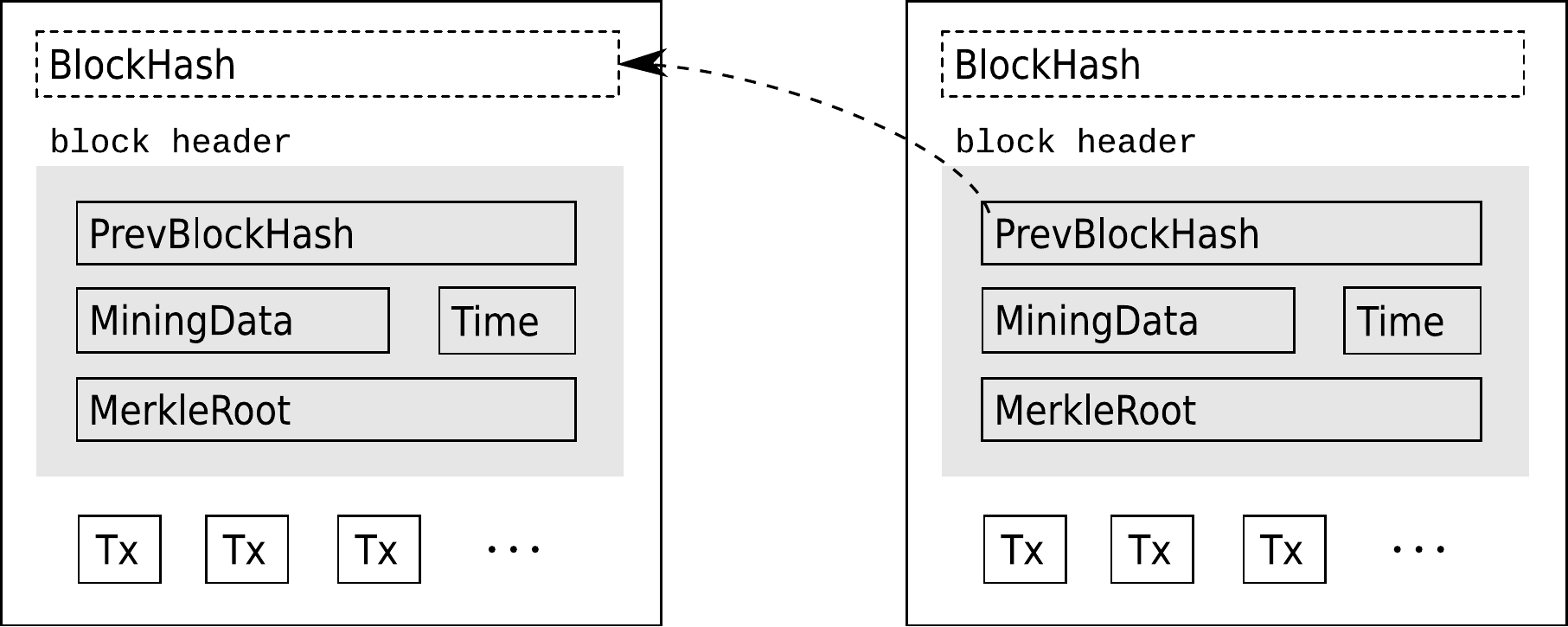}
  \caption{Block data fields.}
  \label{fig:block}
\end{figure}

As will be of little surprise to the reader, blockchains are built of
\emph{blocks}, as also depicted in \cref{fig:block}.
We model a block as consisting of one \emph{block header} and a set of transactions \texttt{Tx}.
\emph{Block hashes} are formed by cryptographically hashing the content of
block headers.
Block headers include the hash of the preceding block in the chain
(\texttt{PrevBlockHash}), a cryptographic commitment to the set of transactions
in the block (typically the root of a Merkle tree formed from transaction
hashes, hence \texttt{MerkleRoot}), a timestamp (\texttt{Time}) and different
types of \texttt{MiningData} depending on the used consensus algorithm.
In the case of a proof-of-work system, for example, this abstract data field
would include a nonce found by miners. In a permissioned system, a
miner's signature might be included.

\begin{figure}[htpb]
  \centering
  \includegraphics[width=0.8\linewidth]{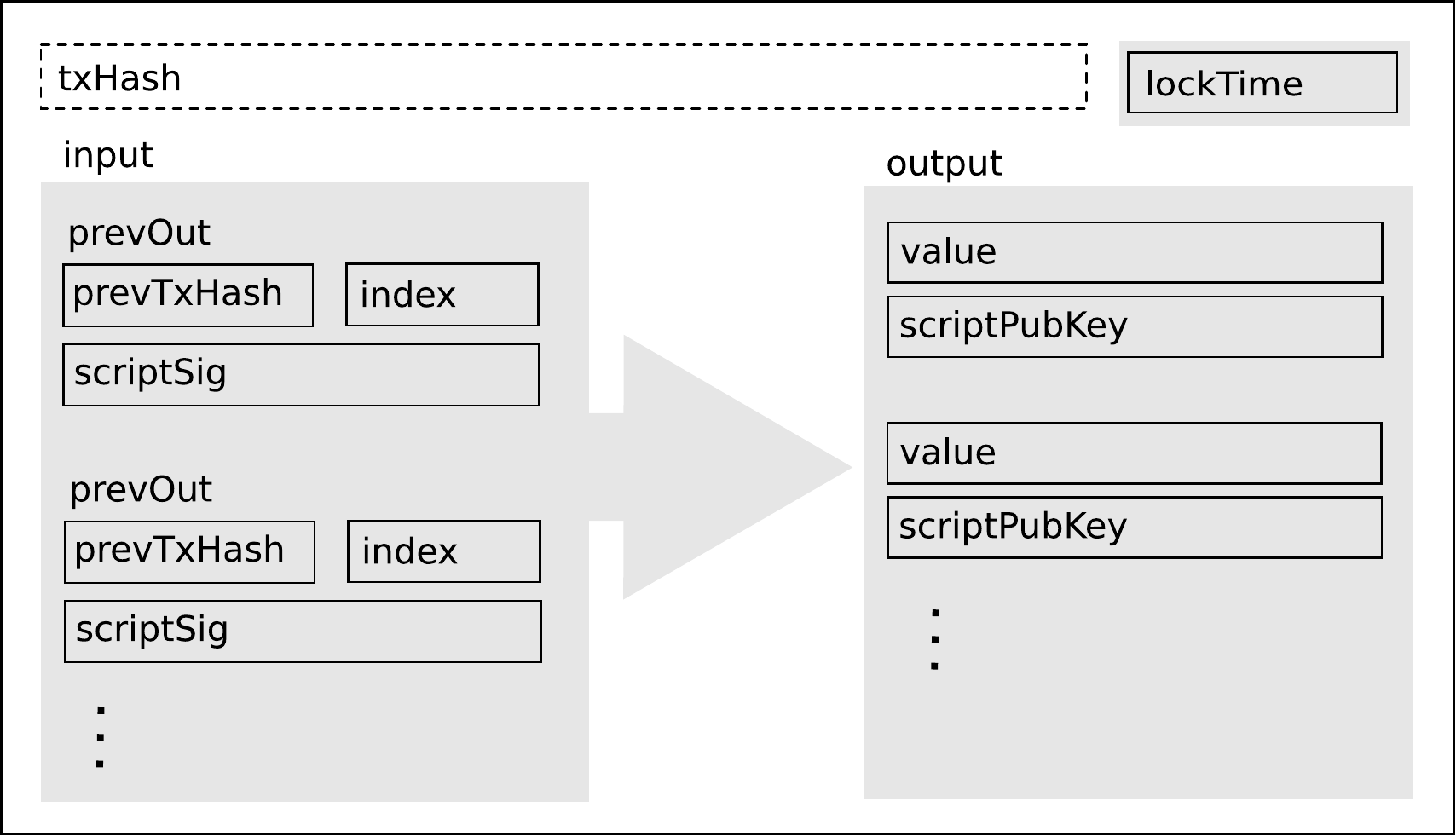}
  \caption{Transaction data fields.}
  \label{fig:transaction}
\end{figure}

The model we used for transactions is depicted in \cref{fig:transaction}.
Transactions are typically identified by a
\emph{transaction identifier}~(\emph{TXID}) identical to a cryptographic hash of the
transaction, i.\,e., \texttt{txHash}.
Transactions can include a \texttt{lockTime} value that encodes the earliest
time that the transaction is allowed to be included in a block.
The main part of most transactions is, however, comprised of their
\emph{inputs} and \emph{outputs}.
Outputs encode \texttt{value} units of cryptocurrency that can be spent, with
a \texttt{scriptPubKey} encoding prerequisites for doing so (typically based on
forming a correct signature with a predefined public key).
Inputs include a reference to outputs of previous transactions and include
a solution to output scripts via the \texttt{scriptSig} field.
Transaction outputs can be used in such a way only once,
allowing them to be differentiated into \emph{spent} and \emph{unspent}
transaction outputs, \emph{STXOs} and \emph{UTXOs}.

Note that for simplicity, our model does not explicitly consider recent developments
such as \emph{Segregated Witness}~(\emph{SegWit})~\cite{lombrozo2015bip141segwit}.
We argue that this is without loss of generality.
For example, while SegWit introduces a new \emph{witness} structure holding
data complementary to \texttt{scriptSig} but stored outside of transactions,
the witness structure can also be modelled as a part of \texttt{scriptSig}
without distorting the general discussion of FPLE.
In \cref{sec:poc}, we focus more on SegWit specifically when discussing
challenges related to the practical implementation of FPLE.

We assume that
blocks are validated sequentially starting from the
genesis block (until reaching the current blockchain "tip"). Validation includes checks
such as:
\begin{itemize}
  \item Is \texttt{PrevBlockHash} the hash of the previous block?
  \item Taking \texttt{MiningData} into account, have the mining prerequisites\footnote{
      Such as the current difficulty target, when using proof-of-work.
    } been met?
  \item Does \texttt{MerkleRoot} match the set of transactions?
  \item Are all included transactions valid?
\end{itemize}

The validation of transactions is based on checks such as:
\begin{itemize}
  \item Do all input scripts satisfy the output scripts they reference?
  \item Are all referenced outputs unspent?
  \item Do the \texttt{value} fields add up correctly\footnote{
      Also taking current \emph{fee} rates into account.
    }?
  \item Has \texttt{lockTime} passed already?
\end{itemize}

For simplicity, we assume that once a block has been processed and deemed
correct, it is not validated again.
Blockchains are per design append-only,
with the validation conditions of a block depending only on data included in previous blocks or the block itself.
Consequently, data which won't be needed for validating subsequent blocks can
be safely and trivially erased, or \emph{pruned}, after it has been validated
and enough blocks have been included on top of if to make a successful fork
preceding the block unlikely (s.a. \cref{sub:pruning}).
In our model as well as in popular networks such as Bitcoin, such
\emph{non-future-relevant} data fields include raw block data (after
future-relevant data items have been stored in other forms such as an UTXO
database), STXOs and provably unspendable outputs\footnote{
  Outputs with a \texttt{scriptPubKey} that always returns \texttt{false},
  for all possible \texttt{scriptSig}.
}.
On the other hand, \emph{future-relevant} data fields, i.\,e., data potentially needed
for the validation of yet unprocessed (or unmined) blocks, include block
hashes (relevant, e.\,g., for the immediately following block),
UTXOs (including their respective transaction IDs) and,
in general, lock times.

A central contribution of our paper is the proposal and functional evaluation
of a backwards-compatible approach to the safe erasure of
future-relevant data fields.
We base the subsequent discussion on the challenge of erasing transaction
parts, with a special focus on the erasure of
data stored in output scripts (\texttt{scriptPubKey}).
Erasing from output scripts is both challenging,
as they are future-relevant, and highly relevant,
as output scripts are the most commonly used storage location for
arbitrary data in Bitcoin~\cite{matzutt2018quantitative} (and likely other networks).
The in-depth discussion of erasure from other future-relevant fields, such as
\texttt{value}, \texttt{lockTime} and \texttt{MiningData}, as well as from
parts of transaction and block hashes (where data could be inserted using brute-force methods),
exceeds the scope of this paper.
The listed fields are currently less relevant in practice and local erasure from them can
conceivably be realized with only minor adaptations to the strategies we introduce here.

\subsection{Erasing Transaction Parts}
\label{sub:modus_erase}

Our main use case are scenarios in which
the desire to erase part of a transaction, such as a transaction output, arises some time
after the respective transaction
has been included on the active chain.
Here, the data is already stored locally by full nodes when the need to erase it arises.
Arbitrary data on blockchains is often not discovered immediately,
but only after careful analysis (such as in \cite{matzutt2018quantitative}) or media reporting.
Data erasure requests as per GDPR (cf. \cref{sub:data_protection}) are another context
in which the need to erase usually arises some time after the data in question
has been included on the blockchain.

With FPLE, the node operator can mark transaction parts as erased whenever deemed necessary by him.
As a result of marking a part $X$ of a locally stored transaction $T$ as erased:
\begin{itemize}
  \item $T$ is stored in the erasure database, with $X$ overwritten by
    substitute values, yielding $T'$.
  \item From its original storage locations, $T$ is physically
    erased or overwritten with $T'$
    in such a way that it cannot be reconstructed.
  \item Operations depending on stored transactions consult the erasure database
    to ensure that relevant input data hasn't been erased. If it has been
    erased, the stored redacted transaction (such as $T'$) is used for subsequent operations.
    This is especially interesting in the case of
    \emph{unspent transaction outputs} (UTXOs).
\end{itemize}

\begin{figure}[htpb]
  \centering
  \begin{tikzpicture}[node distance = 1.5cm, auto]

    \small

    \tikzstyle{block} = [rectangle, fill=gray!22,
    text width=20em, text centered, rounded corners, minimum height=3.5em]
    \tikzstyle{line} = [draw, -latex']

    \node[block] (init) {User input: erase parts $X$ of transaction $T$ with TXID $i_T$};
    \node[block, below of=init] (modify) {Construct $T'$ from $T$ such that $X
      \not\subset T'$};
    \node[block, below of=modify] (store) {Store $T'$ in erasure database with key $i_T$};
    \node[block, below of=store] (remove) {Physically remove $T$ from local storage};

    \path[line] (init) -- (modify);
    \path[line] (modify) -- (store);
    \path[line] (store) -- (remove);
  \end{tikzpicture}
  \caption{Erasure of transaction parts.}
  \label{fig:modus_erase}
\end{figure}
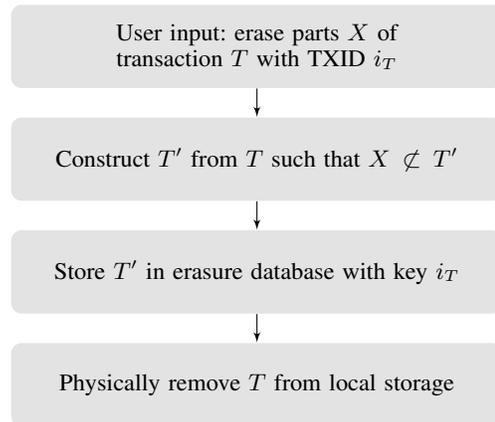

The steps of erasure are also visualized in \cref{fig:modus_erase}.
Conceptually, the erasure database is a key-value store mapping TXIDs to
redacted transactions, i.\,e.,
for the transaction $T$ with TXID $i_T$ it stores the tuple ($i_T$, $T'$).
Additional data, such as the hash of the transaction's block, can be included as well in a practical implementation.

\subsection{Validating with Erased Data}
\label{sub:modus_validate}

In \cref{sub:utxo_model} we discussed the existence of future-relevant data
fields that might be needed for validating later blocks.
This is challenging when erasing, as, e.\,g., spendable UTXOs can easily contain
erasure-worthy data~\cite{matzutt2018quantitative}.
Naively erasing an UTXO opens up the danger that future transactions
(and hence, blocks) will not be deemed correct by the erasing node, leading to a fork.
We propose to pragmatically defuse this challenge
(and avoid the fork) by locally enforcing two simple rules:

\begin{enumerate}
  \item\label{enum:spendable:1} Unconfirmed transactions that reference erased
    data (typically transaction outputs) are always considered invalid and not relayed to other peers.
  \item\label{enum:spendable:2} For confirmed transactions, the SPV heuristic is used for all aspects
    of transactions that cannot be verified due to previously erased data.
    That is, it is assumed that if the transaction was included in a block,
    some miner deemed it to be correct and the transaction is therefore likely
    to remain part of the consensus.
\end{enumerate}

\begin{figure}[htpb]
  \centering
  \begin{tikzpicture}[node distance = 1.5cm, auto]

    \small

    \tikzstyle{decision} = [diamond, draw, fill=gray!22,
    text width=5em, text centered, node distance=3.5cm, inner sep=0pt]
    \tikzstyle{block} = [rectangle, fill=gray!22,
    text width=10em, text centered, rounded corners, minimum height=4em]
    \tikzstyle{sideblock} = [rectangle, fill=gray!22,
    text width=5em, text centered, rounded corners, minimum height=4em]
    \tikzstyle{line} = [draw, -latex']

    \node[block] (init) {Received transaction $T$ with TXID $i_T$ (e.\,g., as part of block)};
    \node[decision, below=0.5cm of init] (erased) {$i_T$ is in erasure database?};
    \node[sideblock, right=1cm of erased] (ignore) {Don't store or process $T$, forget it immediately};
    \node[decision, below of=erased] (depends) {$T$ depends on erased data?};
    \node[sideblock, left=1cm of depends] (normal) {Validate and store $T$ as usual};
    \node[decision, below of=depends] (mined) {$T$ is mined to chain?};
    \node[block, below=0.5cm of mined] (spv) {Assume $T$ is valid, store as usual};

    \path[line] (init) -- (erased);
    \path[line] (erased) -- node[near start] {yes} (ignore);
    \path[line] (erased) -- node[near start] {no} (depends);
    \path[line] (depends) -- node[near start] {no} (normal);
    \path[line] (depends) -- node[near start] {yes} (mined);
    \path[line] (mined) -- node[near start] {yes} (spv);
    \path[line] (mined) -| node[near start] {no} (ignore);
  \end{tikzpicture}
  \caption{Validation of incoming transactions.}
  \label{fig:modus_validate}
\end{figure}
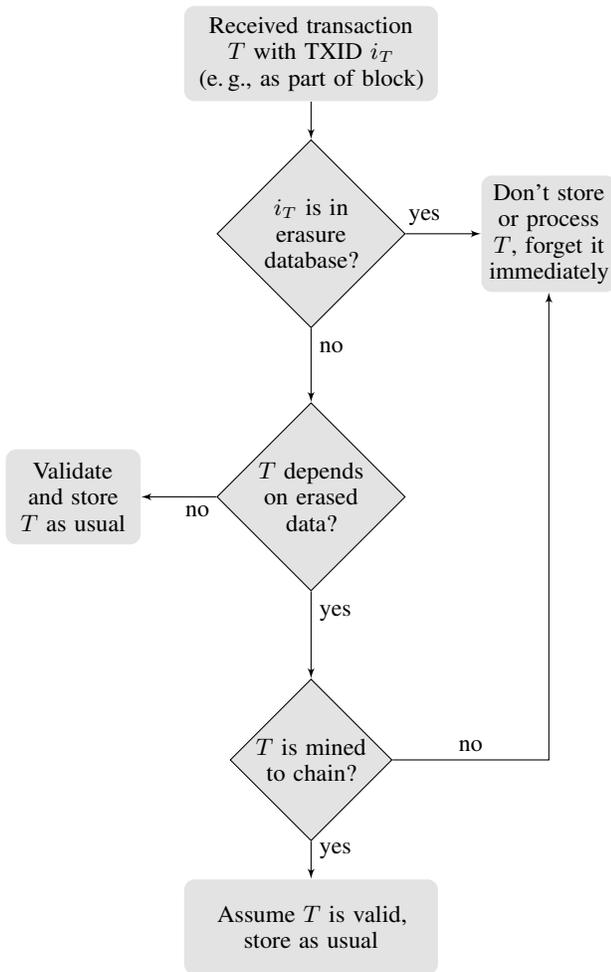

See also \cref{fig:modus_validate} for a visualization of the proposed validation logic.
Note that in various more specific cases, and potentially also in the general case that is in focus here,
the proposed validation logic can be further extended to eliminate the reliance on the SPV heuristic.
We discuss approaches towards this goal in \cref{sub:trustless}.

\subsection{Security Implications of SPV Fallback}
\label{sub:implications}

\begin{figure}[htpb]
  \centering
  \begin{tikzpicture}[node distance = 1.5cm, auto]

    \small

    \tikzstyle{decision} = [diamond, draw, fill=gray!22,
    text width=5em, text centered, node distance=3.5cm, inner sep=0pt]
    \tikzstyle{block} = [rectangle, fill=gray!22,
    text width=18em, text centered, rounded corners, minimum height=4em]
    \tikzstyle{sideblock} = [rectangle, fill=gray!22,
    text width=8em, text centered, rounded corners, minimum height=4em]
    \tikzstyle{line} = [draw, -latex']

    \node[block] (init) {Transaction $T_s$ spends from $T_e$,
        but spend is only valid if $T_e$ has been erased};
    \node[decision, below=0.5cm of init] (mined) {$T_s$ is mined to chain?};
    \node[sideblock, right=1cm of mined] (ignore) {Every honest node discards $T_s$};
    \node[decision, below=0.5cm of mined] (majority) {A deciding majority erased $T_e$?};
    \node[sideblock, right=1cm of majority] (global) {The network accepts that $T_s$ is valid};
    \node[block, below=0.5cm of majority] (fouled) {Nodes that erased $T_e$ consider $T_s$ valid as long as it's part of the longest chain};

    \path[line] (init) -- (mined);
    \path[line] (mined) -- node[near start] {no} (ignore);
    \path[line] (mined) -- node[near start] {yes} (majority);
    \path[line] (majority) -- node[near start] {yes} (global);
    \path[line] (majority) -- node[near start] {no} (fouled);
  \end{tikzpicture}
  \caption{Implications of spending from erased outputs.}
  \label{fig:implications}
\end{figure}
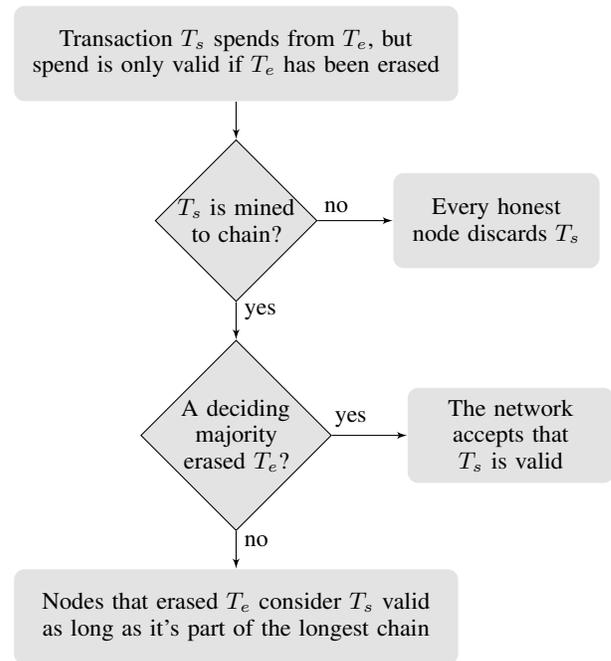

We will now discuss the implications of FPLE for the security of the
erasing node and the overall blockchain system.
Specifically, the rules proposed in \cref{sub:modus_validate} enable a scenario where
a transaction $T_s$ is considered invalid by
non-erasing nodes but valid by some erasing nodes.
In the following, we assume that $T_s$
attempts to spend from an output of transaction $T_e$ with an invalid input
script, and that some nodes erase precisely the output of $T_e$ that is
referenced by $T_s$'s invalid input script.
See also \cref{fig:implications} for a diagram-based representation of this
scenario and its implications.

The described scenario becomes controversial \emph{only} in the case that some miner
included the ("invalid") $T_s$ in a block $B_s$.
Note that due to rule~\ref{enum:spendable:1},
$B_s$ can be mined only by a dishonest or non-validating miner.
As we will see shortly, the miner of $B_s$ risks losing his block
reward as the fork he causes is unlikely to succeed.

Should such a block $B_s$ nevertheless be mined, the implications depend on whether a
\emph{deciding majority} of nodes, i.\,e., a group of miners with
sufficient consensus weight to unilaterally perpetuate the longest chain,
have also erased the relevant transaction output of $T_e$.

\paragraph{No deciding majority erased the relevant transaction output}
    The problematic block $B_s$ will not become part of the global consensus.
    An erasing node might, however, act on the belief that $T_s$ is valid,
    making is susceptible to a double-spend attack on funds locked in erased
    UTXOs.
    Again, a prerequisite for this attack is the mining of an invalid block
    that will be rejected by the deciding majority---a high economic cost.
    The costs for the attack can be increased further by waiting for a number
    of confirmations before trusting the block, a measure that is recommended
    even when running a regular non-erasing full node~\cite{gervais2016security}.

\paragraph{A deciding majority erased the relevant transaction output}
    The crafted spending transaction $T_s$ and the block $B_s$ including it will become
    part of the longest chain.
    The funds locked by the erased data in $T_e$ will effectively become stolen,
    i.\,e., transferred without correct credentials.
    For decentralized systems like Bitcoin, the probability of such a
    scenario taking place in practice are arguably small (off-chain consensus and
    cooperation are notoriously difficult).
    Still, we would argue:
    \begin{itemize}
      \item
        If the UTXO was erased because it contains data whose possession and redistribution
        is illegal or considered unwanted by a deciding majority,
        the recipient of the funds arguably has no claim to being protected by the network.
        Recipients usually provide key input to forming transaction
        outputs (e.\,g., their cryptocurrency address) and are free to choose whether they
        accept a non-standard UTXO as payment or not.
      \item
        If the UTXO was erased because a user rightfully requested it,
        for example when considering requests for the removal of privacy-relevant data (cf. \cref{sub:data_protection}),
        the user is removing his
        consent for the use of that data.
        By doing so, he is also accepting any potential consequences to
        the security of his own funds that result from the erasure.
      \item
        In any case, if a deciding majority considers that an output should
        henceforth be viewed as "anyone-can-spend" (by erasing it locally),
        this is arguably an argument in itself for accepting this view as
        consensus.
    \end{itemize}
    Also note that fund owners are always free to move their funds to a "safer" location, should
    a relevantly broad consensus for an erasure become likely.

\subsection{Towards Trustless Validation}
\label{sub:trustless}

The FPLE approach outlined so far implies a trade-off between the erasure of
some data and the ability to gaplessly validate the complete blockchain and transaction graph.
As pointers for subsequent works,
we will now discuss how erasure could be realized without introducing validation gaps that must be bridged by the SPV heuristic or other forms of trust.

\emph{Tailored erasure mechanisms} are possible in reply to various specific data storage approaches.
In Bitcoin, for example, around \SI{99}{\percent}\footnote{
 Based on the state of the UTXO set in October 2017 \cite{delgado2019utxo}.
}
of UTXOs are based on a \emph{pay-to-hash}
template: their output scripts consist mostly of a cryptographic hash value $h$.
In order for a spend to be successful,
some part $X_s$ of the provided \texttt{scriptSig}
(such as a public key or a piece of script code) must, upon hashing, result in
$h$, i.\,e., the spend is only valid if $\hash(X_s) = h$.
If $h$ has been erased, this check cannot be performed and an SPV approach
must be used.
However, an integrated implementation can also submit $h$ to a second round of
hashing before erasing it, storing the resulting $h' := \hash(h)$ in its erasure database.
Being the output of a cryptographic hash function, $h'$ is unlikely to contain problematic data
(the likelihood can be further reduced by using a salt value during hashing).
It is furthermore unfeasible to reconstruct $h$ with only knowledge of $h'$.
Having stored $h'$ for each erased $h$, the local validation rules can now be
extended to also submit the relevant parts of proposed input scripts to a
second round of hashing, comparing the result with the stored $h'$.
If testing that $\hash(X_s) = h$ is not possible due to an erased $h$, the
erasing node can instead check whether $\hash(\hash(X_s)) = h'$.
This allows for a fully independent and "\emph{trustless}" validation, i.\,e.,
without resorting to the SPV heuristic.
While a similar strategy can also be applied in other scenarios
(e.\,g., when considering data encoded in the last bits of transaction or block hashes),
the specifics depend heavily on the individual validation context.

A more \emph{universally applicable approach} towards trustless validation with erased data is conceivable as well.
Before erasure, impacted data could be transformed using an appropriate
\emph{homomorphic encryption} scheme so that its reconstruction is impossible
while relevant validation operations can still be performed in a trustless
manner (by analogously encrypting the input data and completing the operation
within the homomorphic system). We leave the thorough investigation of this
promising (albeit more complex) approach to follow-up works.

\subsection{(Not) Receiving Data and Bootstrapping New Nodes}
\label{sub:bootstrapping}

So far we focused on the case that the desire to erase a chunk of data arises
some time after it has been received, stored and locally validated.
Once some data has been processed and has been marked as erased, there is no need to request it again from other nodes.
However, a node operator might be aware of problematic data in some
block or transaction that he doesn't yet store, e.\,g., when bootstrapping a new
node that still hasn't synced to the rest of the network.
We see two solutions to enabling FPLE in this scenario---accepting the storage
for a short duration, and obtaining relevant erasure database entries from a trusted party.

In practice, depending on the reasons for erasure,
it might be acceptable to willingly \emph{request and store problematic data
for a very short duration}, with the sole goal of validating it once and
then discarding it.
Since the receipt, processing and erasure of the data is automatized,
with no possibility for the node operator to later extract the problematic
data, such an approach might be applicable
even in cases where the legal obligation for erasure exists.
Clearly, the applicability of this reasoning is highly dependent on the individual
case.

Should even the receipt of certain transactions or blocks
be undesired (or impossible, e.g., if all reachable nodes already erased them),
a remaining solution is to
\emph{obtain erasure database entries from a trusted party}.
As discussed in \cref{sub:modus_erase}
for the case of erasing transaction outputs, an erasure database
entry consists of the TXID $i_T$ of a transaction $T$ containing problematic data
and a modified version of that transaction $T'$ that doesn't include the
problematic data, i.\,e., the tuple $(i_T, T')$.
Notably, the trusted source of such entries must be trusted to leave non-problematic outputs
of $T$ in their original form when constructing $T'$, as otherwise the
validation challenges discussed in \cref{sub:modus_validate} will unnecessarily
be extended to them.
In some scenarios, such as when strong legal obligations for erasure exists
(e.\,g., when the rights of the child are threatened; cf. \cref{sub:criminal_law}),
it is conceivable that a well regulated public institution performs the role of
trusted erasure database source, reducing the practical risks
of erasure without validation.

\section{Proof-of-Concept Application to Bitcoin}
\label{sec:poc}

We now introduce a proof-of-concept implementation\footnote{
  \url{https://github.com/marfl/bitcoind-erase}
}
of FPLE for Bitcoin
targeting current versions\footnote{
  We validated our implementation against version 0.17.1.
} of the \emph{Bitcoin Core} node software
implementation, also known as \emph{bitcoind}.
Our main goal is to demonstrate the practical feasibility of FPLE.
We also hope to provide groundwork for the development of a more general
erasure tool for node operators.

\subsection{Overview and Scope}
\label{sub:poc_overview}

Our proposal for implementing FPLE for production use is to modify existing
node software such as bitcoind,
extending it with an erasure database and suitable hooks in the validation process.
For our proof-of-concept, we explored an alternative, less invasive
instantiation of FPLE that enables the erasure of already stored transaction
outputs without requiring changes to the bitcoind binary.
We chose this approach because
it allows for a quicker validation of our general FPLE proposal.

Our tool works by modifying existing data stores of bitcoind in such a way
that selected transaction outputs are (1) pruned from the raw block storage
and (2) their \texttt{scriptPubKey}s are overwritten in the local UTXO database to
the equivalent of "any input can spend".
Selected outputs' original contents are erased while the local node stays in
sync with the network even if any of the outputs becomes spent.

A major constraint of our current implementation is its reliance on existing
\emph{pruning} logic for erasing from bitcoind's raw block store.
Building upon bitcoind's built-in pruning functionality leads
to the following specific constraints:
\begin{itemize}
  \item More data is erased than required as bitcoind's pruning
    functionality erases all blocks up to a target height.
    Our tool is therefore currently usable only when full blockchain archival
    and indexing is not a requirement.
    Notably however, nodes still validate the whole blockchain.
  \item Full erasure of an output can happen only after the node is fully
    synced up to the point where the output is "buried" under 300 blocks,
    i.\,e. not sooner than roughly 50 hours after content has been included on
    the active chain.
\end{itemize}

Additionally, since we can't guarantee that transactions spending from erased
outputs are ignored when building blocks, our current implementation should
not be used with mining nodes.
Lastly, our tool can currently safely erase only transaction outputs
that are not SegWit outputs.
For supporting the erasure of SegWit outputs, modifications of bitcoind are
currently unavoidable.
More background on SegWit-support and other implementation details is given in
\cref{sub:implementation_details}.

\subsection{Proof-of-Concept in Action}
\label{sub:functional_evaluation}

Introducing our tool's functionality through an example,
we will now describe how we erased transaction
\texttt{c206e}\footnote{Full TXID:
  \href{https://www.blockchain.com/btc/tx/c206e8fff656f07b27dac831ef9b956792bae4e76a2cb43f14f49f0298bf2c2f}{\texttt{c206e8ff\-f656f07b\-27dac831\-ef9b9567\-92bae4\-e7\-6a\-2c\-b4\-3f\-14f49f02\-98bf2c2f}.}
}.
We also erased several other transactions from our node that we discovered during our research
and consider undesirable to store and distribute.
We choose not to discuss these transactions here due to ethical and legal considerations.
However, we would argue that with its capability for erasure, and since we
already erased all problematic transactions known to us, our node is the first
full Bitcoin node that is free from existential legal risks
related to the storage of potentially problematic data.

Transaction \texttt{c206e} is included on the active mainnet chain
in block
\texttt{892a0}\footnote{Full block hash:
  \href{https://www.blockchain.com/btc/block-index/1189978}{\texttt{00000000\-00000000\-0569\-4b14\-df70\-b8\-db\-db\-8e\-0c\-92\-234f5d30\-209781c0\-941892a0}.}
}.
It has 155 outputs, two of which were spent at the time of writing.
When interpreted as a JPEG, the outputs' payload combines to
a humorous image of a young muscular man wearing a face-covering gas mask and sunglasses.
When interpreted as text, the transactions' outputs furthermore contain the
message "Hi mom! I love you.".

We started syncing our node
from the genesis block, and shortly after receiving block \texttt{892a0}
paused our node and marked \texttt{c206e}'s outputs for erasure,
to simulate the case that the transaction has just been included on chain and
none of its outputs are spent.
We marked all 155 outputs of \texttt{c206e} for erasure.
The relevant portion of our tool's configuration file is provided
below (edited for brevity):
\begin{lstlisting}[language=java]
{
  "bitcoind_dir" : "~/.bitcoin/",
  "chain" : "mainnet",
  "erase" : {
    "00000...892a0" : {
      "c206e...f2c2f" : [0, 1, 2, ..., 153, 154]
    },
    ...
  }
}
\end{lstlisting}

All transaction outputs of \texttt{c206e} follow the \emph{Pay-to-PubKeyHash}~(\emph{P2PKH})
template and therefore have an output script of the form:
\begin{lstlisting}
scriptPubKey: OP_DUP OP_HASH160 <X>
  OP_EQUALVERIFY OP_CHECKSIG
\end{lstlisting}

With \texttt{X} being used here as a placeholder for
20 bytes of data (different for each output). Typically, \texttt{X} is the hash of the public
key that needs to be provided in order to spend the output. For most
outputs in \texttt{c206e}, however, it is likely that since they
contain meaningful JPEG data, no public key is currently known that
is able to hash to their payload and therefore would enable spending the
associated outputs.
While theoretically spendable, transaction outputs containing arbitrary data
are therefore very unlikely to be spent in the foreseeable future.
The observation that this is commonly the case when arbitrary data is included in
transaction outputs motivated our pragmatic "any input can spend" replacement
approach. Implemented in our proof-of-concept, transaction outputs marked in the
configuration (here: all transaction outputs of \texttt{c206e}) are modified
so that their \texttt{scriptPubKey} becomes of the form:
\begin{lstlisting}
scriptPubKey: OP_TRUE
\end{lstlisting}

The \texttt{TRUE} opcode pushes a 1 to the stack. Since it is the only opcode
in the output script, any \texttt{scriptSig} that doesn't deliberately add termination
conditions (regular input scripts don't) will lead to a positive evaluation
and therefore the ability to spend.
We can confirm that rewriting outputs in this way effectively prevents cases in
which the local node becomes out-of-sync because it can't validate mined
transactions spending from erased outputs.
We validated the claim using
automated regression tests described in \cref{sub:implementation_details}
as well as through manual mainnet experiments like the one described here with
transaction \texttt{c206e}.
Following the erasure of \texttt{c206e}'s outputs, we resumed our node, which then
successfully synced to the network (despite two spends from erased outputs of
\texttt{c206e}). At the time of writing,
our node has been running without intervention for more than 2 months,
remaining in sync with the network and fully validating all incoming blocks.

\subsection{Implementation Details}
\label{sub:implementation_details}

We now touch upon several potentially interesting implementation details
related to bitcoind storage locations, automated tests and SegWit support.

\subsubsection{Data storage locations}
\label{ssub:blockchain_data_storage_locations}

According to our analysis of its code base and data folder structure,
bitcoind stores potentially relevant
blockchain data in the following locations (relative to its
data folder):

\begin{itemize}
  \item \texttt{blocks/}: Raw block data stored in a custom data format.
    Our tool erases from this location using bitcoind's built-in pruning functionality.
  \item \texttt{chainstate/}: LevelDB database storing different aspects of
    the current blockchain state. Most importantly,
    this \emph{chainstate database} holds a full copy of the current UTXO set,
    i.\,e., all current UTXOs.
    It is in this database that our tool erases the payload of UTXOs marked
    for erasure, by overwriting their \texttt{scriptPubKey}.
  \item \texttt{indexes/txindex/}: An optional transaction index database
    that is not maintained when pruning is enabled and we therefore ignored in
    our implementation.
  \item \texttt{mempool.dat}: A collection of transactions that are
    not yet included on the active chain but are valid and likely
    to be included soon.
    We ignore this data store here, focusing instead on the
    use-case of erasing data after it has been included on the active
    chain.
  \item Wallet-related files and logs, which we also ignore because
    they are trivially erased if necessary.
\end{itemize}

\subsubsection{Validation through automated tests}
\label{ssub:validation_through_automated_tests}

We developed an extensive test suite that leverages
bitcoind's regression testing mode.
Among other things, we automatically validate that:

\begin{itemize}
  \item Blocks containing erased transactions are not stored in the raw blocks
    storage, are not requested from peers and are not shared with peers.
  \item Erased transaction outputs are either not stored in the chainstate database
    or stored in the aforementioned modified form.
  \item Transactions with erased outputs are also not obtainable via
    bitcoind's RPC API.
  \item Mined blocks containing transactions that spend from erased outputs
    are regularly accepted.
\end{itemize}

\subsubsection{SegWit support}
\label{ssub:segwit_support}

\emph{Segregated Witness} (\emph{SegWit})~\cite{lombrozo2015bip141segwit}
is a recent change to the Bitcoin protocol
introducing a new \emph{witness} structure that, for some transaction
outputs, needs to be provided in addition to the \texttt{scriptSig}.
While the change is orthogonal to our general FPLE approach,
it has direct implications for our current proof-of-concept implementation.
In bitcoind version 0.17.1 (and possibly others),
if witness data was provided for spending an output, it must be used during
the validation. Otherwise the validation fails.
When spending from SegWit UTXOs,
witness data is provided together with a \texttt{scriptSig}, i.\,e., it can't be
influenced at the time of erasing. Pragmatic workarounds towards ignoring the
script contained in the witness are possible,
e.\,g., by manipulating the SegWit version byte stored in \texttt{scriptPubKey}.
However, with the validation rules currently
implemented in bitcoind, validation also fails if
the spending of such a modified (a now obviously SegWit-based)
output is attempted by a pre-SegWit
(\emph{legacy}) input with nonzero \texttt{scriptSig}.
A different rewriting approach is therefore needed for erasing SegWit
outputs than for erasing pre-SegWit outputs.
Unfortunately, a commonly used template is to nest
SegWit-spendable outputs in a (legacy) \emph{Pay-to-Script-Hash} (\emph{P2SH}) output,
which makes them indistinguishable from legacy P2SH transaction outputs.
SegWit-related challenges can easily be surmounted when implementing FPLE by
directly modifying the validation rules in bitcoind.
Motivated by our own erasure goals,
our non-invasive proof-of-concept focuses on the safe erasure
of outputs that are clearly not SegWit-based, such as P2PKH outputs.

\section{Why would Someone Want to Erase?}
\label{sec:why_erase_}

In the following, we give an in-depth motivation of why the availability
of a pragmatic local erasure solution is highly
desirable for node operators in practice.
An overview over potential negative implications of arbitrary data storage on
public blockchains like Bitcoin can be found
in~\cite{matzutt2018quantitative}.
There, Matzutt et al. conclude that participation in blockchain-based systems could be
considered illegal once illegal content has been included on-chain.
Based on our own legal analysis, however, we argue that liability can be avoided if no problematic data is actually stored or distributed, which can be achieved at the node-individual level using FPLE.

We explain our reasoning at the example of criminal law with respect to the
storage and distribution of problematic images.
We then briefly touch upon data protection law
and
point out other legal and non-legal reasons for the desirability of erasure.
While our specific legal arguments are rooted in EU law
they are likely transferable to various similar jurisdictions.

\subsection{Criminal Law}
\label{sub:criminal_law}

According to the prevailing narrative,
full nodes store the whole blockchain, which
is public, immutable and cannot be erased.
The implication is that they potentially
also store and distribute data of which the storage and distribution is illegal.
However, criminal liability could be avoided if immediate deletion of such data was
possible.

The question of deleting data from the blockchain gains urgent practical significance as images containing juvenile pornographic content might already be included on the Bitcoin blockchain~\cite{matzutt2018quantitative}.
Additionally,
no possibility is apparent for preventing people from adding illegal content to the Bitcoin blockchain in the future (cf. \cref{sec:related_work}).

The possession and distribution of child and juvenile pornography is subject to criminal liability in most jurisdictions.
Although, to the best of our knowledge, no court has decided on the liability of full node operators for images stored on the blockchain yet, legal grounds for operating a Bitcoin full node are precarious.

A possible legal solution to the matter can be found within EU law.
It might be possible to apply Art. 14 of the \emph{E-Commerce Directive}~\cite{eu2000ecommerce} to blockchain full nodes.
The article was drafted for \emph{information society services}, services such as hosting providers that store information originally provided by recipients of the service.
\cite[Art. 14]{eu2000ecommerce} provides a safe harbour for such services.
Pursuant to this legal norm, the provider of the service is not liable for the information stored at the request of a recipient of the service, on condition that:
\qq{(a) the provider does not have actual knowledge of illegal activity or information […]}
or
\qq{(b) the provider, upon obtaining such knowledge or awareness, acts expeditiously to remove or to disable access to the information}.

Whether the safe harbour of Art. 14 E-Commerce Directive is available for blockchain full nodes is under legal debate~\cite{beaucamp2018strafbarkeit}, with no court rulings known to the authors.
Provided that \cite[Art. 14]{eu2000ecommerce} can be applied to blockchain full nodes, the argument can be made that the operator of the blockchain full node can only be held criminally liable if he has actual knowledge of the possession and distribution of the incriminating image.
Whether actual knowledge can be assumed depends on individual circumstances.
However, also under \cite[Art. 14]{eu2000ecommerce}, the node operator is obliged to erase the relevant data once they are informed of its existence in order to avoid criminal liability.
Only disabling access to data containing child and juvenile pornography, without physically erasing it, would be insufficient to comply with the law as knowing possession remains a criminal offence.

\subsection{Privacy and Data Protection}
\label{sub:data_protection}

As also mentioned in \cref{sub:filtering_data},
achieving strong privacy-guarantees for data included on a public blockchain is
non-trivial from a technical standpoint.
If no explicit precautions are taken,
which appears to be the norm~\cite{moser2017anonymous},
confidentiality of transactions is next to not given~\cite{conti2018survey, herrera2016privacy}.
The issue is further compounded when considering non-financial use-cases or
the deliberate posting of privacy-relevant data without consent, e.\,g., as part
of a doxing.
Contemporary data protection standards demand that network participants can also erase data already shared with them,
a requirement that is at odds with the notorious "impossibility" of erasure from blockchains~\cite{berberich2016blockchain}.

Pursuant to Art. 17 of the \emph{General Data Protection Regulation} (GDPR)~\cite{eu2016gdpr} every data subject has the \emph{right to be forgotten},
meaning the right to obtain from the data controller
the erasure of personal data concerning him or her without undue
delay~\cite[Art. 17]{eu2016gdpr}.
This right is subject to certain prerequisites and certain exceptions (see \cite[Art. 17]{eu2016gdpr} for details).
According to the prevailing opinion, compliance with the duty to erase is
given if the information embodied in the data cannot be recovered without
disproportionate effort~\cite{kamann2017dsgvo}.

Whether blockchain full node operators can be defined as data controllers is, however, legally contested~\cite{pesch2017distributed, berberich2016blockchain}.
The realization that local erasure is possible, however, contributes significantly to the legal discussion on who the data controller in a blockchain system is and who can therefore be made responsible for deletion.

The right to be forgotten is the most striking deletion-related conflict point between common blockchain protocols and the GDPR.
However, other norms of the GDPR, for example the necessity of consent for the processing of data \cite[Art. 6]{eu2016gdpr},
could also be in conflict with the features of blockchain full nodes,
leading to more potential reasons why full nodes might need to erase personal data stored on the blockchain.

\subsection{Further Legal, Ethical and Social Norms}
\label{sub:further_norms}

The criminal law and data protection norms mentioned above are only a small selection of legal reasons for the importance of the possibility of at least local deletion of content from a blockchain full node.
Similar legal problems exist concerning intellectual property rights
(where host providers need to be able to respond to takedown requests),
defamation, and malware distribution, to name just a few examples (an overview can also be found in \cite{matzutt2018quantitative}).

Popular public networks like Bitcoin extend worldwide, spanning over a multitude of jurisdictions, but also over diverse individuals subject to different
ethical, religious and social norms.
Local erasure is therefore not only necessary for legal reasons,
but also desirable
for full node operators to avoid storing and distributing content that is in
conflict with their individual norms and values,
even when no network-wide consensus on norms and values exists.

\section{Discussion}
\label{sec:discussion}

In the following, we discuss difficult-to-measure
implications of FPLE for existing blockchain networks.

\subsection{Network Resilience}
\label{sub:network-wide}

As observed in \cref{sec:related_work},
it is likely impossible to completely
restrict the encoding of
arbitrary data on a public blockchain like Bitcoin, and that it is
therefore unavoidable that objectionable data (e.\,g., child pornography,
sensitive personal data) will at some point be included, be it
deliberately or by accident.
For example,
malicious actors might deliberately include illegal content
on a blockchain, with the goal of causing legal insecurity
for node operators and thus damaging the
network~\cite{matzutt2018quantitative}.
FPLE protects against this risk: node
operators can erase problematic data locally and in this way remain
compliant, even if the data is still formally "on the chain" and perhaps
stored on other nodes in the network.

Reducing individual legal risks
is central for maintaining large numbers of full nodes and therefore a
healthy blockchain network and ecosystem.
Even more so in situations where compliance is pursued and the identity of node operators is well known,
as when considering businesses depending on full node functionality (exchanges, online wallets)
or federated systems like
\emph{Stellar}~\cite{mazieres2015stellar}.

\subsection{Global Erasure}
\label{sub:global_erasing}

In this paper, we focus on enabling local erasure.
Our main argument for the desirability of a local erasure possibility is that
reasons for erasure are highly individual, making global consensus on this
topic difficult and the enforcement of a unified global policy potentially undesirable.

In cases where a global consensus about the erasure of a given data chunk does exist,
FPLE can also
scale to global erasure as individual node operators implement similar erasure decisions.
We argue that without such off-chain consensus, no effective global erasure is
possible with whatever technical means.
Individual nodes always retain the possibility to keep data that they
already store (possibly in secret), and can't, in general, prove they have forgotten it.

\subsection{Censorship}
\label{sub:censorship}

As a potential downside, the possibility for local erasure could be viewed as
an enabler for censorship in oppressive regimes,
as forcing the erasure of blockchain data becomes decoupled from the
collateral damage of having to altogether forbid the operation of a node.
We believe that the increased flexibility for node operators outweighs such risks
and note that the discussion is similar to considering the tainting of
cryptocurrency-denominated crime proceeds~\cite{anderson2019bitcoin}.

\section{Conclusion and Outlook}
\label{sec:outlook}

In this paper, we question the common narrative that erasure is not possible
for node operators in existing blockchain networks like Bitcoin.
In contrast to existing erasure approaches attempting to erase data globally from all nodes,
we propose a pragmatic local erasure solution.
Our functionality-preserving local erasure (FPLE) approach
empowers node operators to
remove problematic transaction parts from local storage,
with minimal impact to
their capacity to support the network and autonomously validate further transactions.
Challenging implications of erasure, such as the potential inability to
validate some new transactions, are diffused by a set of simple rules that
weaken security guarantees only for transactions directly referencing erased data.
We demonstrate the non-invasive applicability of our solution to existing
protocols using a proof-of-concept implementation for Bitcoin.
We argue that by enabling local erasure, we enable blockchain networks to
embrace a larger range of legal norms and individual values and to therefore truly
become global networks.

Building upon the proposed approach,
possible next steps include the investigation of fully trustless approaches towards FPLE,
e.\,g., based on homomorphic encryption.
We will also further validate the application of
FPLE to other blockchain architectures, towards enabling local erasure
in further popular networks such as Ethereum and Stellar.

\bibliographystyle{IEEEtranS}
\bibliography{paper}

\end{document}